\documentstyle[12pt]{article} \topmargin=-5mm \oddsidemargin=6mm
\newcommand{\cftnote} {\renewcommand{\thefootnote}{\fnsymbol{footnote}}}
\newcommand{\resetftnote}{\setcounter{footnote}{0}}

\newcommand{\be}{\begin{equation}} \newcommand{\ee}{\end{equation}}
\newcommand{\bea}{\begin{eqnarray}} \newcommand{\eea}{\end{eqnarray}}

\def\bm#1{\mbox{\boldmath{$#1$}}}
\def\l{\bm\lambda} \def\a{\bm\alpha} \def\r{\bm\rho}
\def\o{\bm\omega} \def\t{\bm\theta} \def\e{\bm\epsilon} \def\m{\bm\mu} 
\def\W{{\cal W}}

\def\IR{\relax{\rm I\kern-.18em R}}

\begin{document}

\begin{flushright} IMAFF-95-17\\ hep-th/9602001\\ Jan 1996 
\end{flushright}

\cftnote

\vspace{1mm} {\Large \bf \begin{center} 
Landau-Ginzburg Lagrangians of minimal $W$-models with an integrable
perturbation%
\footnote{Research supported in part by the CICYT (Spain) under project
PB92-1092.} \end{center} } \vspace{2mm} 
\setcounter{footnote}{3}
\begin{center} 
{\large Jos{\'e} Gaite%
\footnote{E-mail address: gaite@sisifo.imaff.csic.es}} \\[4mm] 
{\it Instituto de Matem{\'a}ticas y F{\'\i}sica Fundamental, C.S.I.C.,\\ Serrano 123,
28006 Madrid, Spain.}\\ 
\end{center} 
\resetftnote

\vspace{2mm}

\begin{abstract} 
We construct Landau-Ginzburg Lagrangians for minimal bosonic ($N=0$)
$W$-models perturbed with the 
least relevant field, inspired by the theory of $N=2$
supersymmetric Landau-Ginzburg Lagrangians. They agree with the
Lagrangians for unperturbed models previously found with Zamolodchikov's
method. We briefly study their properties, e.g. the perturbation algebra
and the soliton structure. We conclude that the known
properties of $N=2$ solitons (BPS, lines in $W$ plane, etc.) hold as
well. Hence, a connection with a generalized supersymmetric structure
of minimal $W$-models is conjectured.
\end{abstract}

\global\parskip 4pt

\vspace{1cm}


\newpage

\section{Introduction}

The construction of Landau-Ginzburg Lagrangians for 2d conformal field
theories (CFT) is a powerful tool in the study of these theories as a
description of the critical behaviour of statistical models.
Besides, the case with $N=2$ supersymmetry has found wide
application in string theory. It has been shown before that
Landau-Ginzburg Lagrangians can be constructed for a wide class of 2d
CFT, the minimal models of W-algebras \cite{I}. Those Landau-Ginzburg
(LG) Lagrangians describe the multi-critical behaviour of
two-dimensional systems with dihedral $D_n$ symmetry; in particular,
they describe the multi-critical behaviour of the integrable IRF models
introduced by Jimbo {\em et al} \cite{JMO}.

The LG Lagrangians for minimal $W$-models were obtained by a generalization of
the method of Zamolodchikov for the minimal Virasoro models \cite{Zamo}.
Thus a composite field structure was given to the algebra of relevant
primary fields and a field equation ensued from the truncation of this
algebra.  However, the complete Landau potential was obtained only for
the case of $W_3$. For $W_3$-models it was shown that there is a
perturbation that reproduces the ground state structure of the
corresponding IRF model and that this is the maximum possible unfolding
of the potential \cite{I,II}. Although it is possible in principle to
obtain similar results for $n>3$, the algebraic methods utilized for
$n=3$ are hardly generalizable.

The theory of perturbed Landau potentials for 2d CFT with $N=2$
supersymmetry is however well developed \cite{Warner}. When
the perturbation is the least relevant field, one has that the
superpotential is precisely the fusion potential of $SU(n)$ WZW theories
found by Gepner \cite{Gepner}. Furthermore, the $SU(n)$ WZW fusion
algebra coincides with the algebra of chiral fields of the $N=2$ CFT.
Unfortunately, the Gepner potential is complex and hence not suitable for the
non-supersymmetric case. Nevertheless, one can derive from it a bosonic
potential with interesting properties \cite{Fendley et al}.
In the one-variable case ($n=2$) the
$N=2$ bosonic potential has the expected properties for a $N=0$
potential; namely, if we label the minimal models with $p= 3,\ldots$, as
usual, it has degree $2(p-1)$ and $p-1$ minima with zero
energy. However, for $n>2$ its degree is higher than that 
required by the structure of the algebra of relevant fields and moreover
it does 
not constitute a well defined real singularity when the perturbation is
set to zero.
Nevertheless, there exists a remarkable relation between the fusion
algebras of the $SU(n)$ WZW theories and certain field subalgebras of
the $W$-models that indicates that Gepner's construction is also
relevant for $N=0$. We shall rely on it to find perturbed Landau
potentials that agree with those obtained with Zamolodchikov's method.

\section{The algebra of relevant fields of $W$-models}

We will consider only modular 
diagonal invariant models for which we need only spin zero
fields. Therefore, we shall only consider the holomorphic part of
fields. Besides, the underlying Lie algebra of the $W$-symmetry is
always $A_{n-1}$.
The algebra of relevant primary fields of the model $W_{(n)}^p$ has been
described before \cite{I}.
It consists of two parts: First, the fields
of lower dimension, $\Phi(\l\mid\l)$, which were called diagonal, 
with dimension $$\Delta={\l(\l+2\r)\over 2p(p+1)},$$ 
proportional to the first Casimir of the $SU(n)$ representation with
highest weight $\l$. They fill a Weyl alcove of level $k=p-n$.
The second part consists of the fields $$\Phi(\l-\a\mid\l)$$ where
$\a$ is a positive root or, in some cases, the sum of two positive
roots; these fields are called non-diagonal.  They are all
generated successively as powers of the elementary fields 
\bea
x_k= \Phi(\o_k\mid \o_k),~~k=1,\dots,n-1, \\
x_{n-k} = {\bar x}_k,
\eea
with $\o_k$ the $k$th fundamental weight,
according to the standard definition of composite fields \cite{Zamo}.
For the diagonal fields one has
\be
\Phi(\l\mid\l) = \prod_{k=1}^{n-1} x_k^{\mu_k},  \label{diag}
\ee
with $\mu_k$ the Dynkin labels of $\l=\sum \mu_k \o_k$. The subsequent
powers of $x_k$ are identified with non-diagonal fields, though this
identification is not as straightforward as (\ref{diag}) \cite{I}.

There is a distinguished subalgebra of relevant fields, generated by
\be
\epsilon := \Phi(\t\mid \t) = x_1 \bar{x}_1,
\ee
called the thermal subalgebra. Their diagonal fields are
\be
\epsilon^l = \Phi(l\t\mid l\t),~~2\,l\leq p-n,
\ee
and their non-diagonal fields
\be
\epsilon^l = \Phi((p-n-l)\t\mid (p-n-l+1)\t),~~p-n<2\,l\leq 2(p-n).
\ee
The least relevant field of the thermal subalgebra,
\be
\epsilon^{p-n} = \Phi({\bf 0}\mid \t), 
~~\Delta=1-{n\over {p+1}}\stackrel{<}{\sim}1,
\ee
produces the field equation upon multiplication by $x_1$
\be
x_1\,(x_1 \bar{x}_1)^{p-n} = \partial^2 x_1.  \label{fe} 
\ee
However, the ensuing Lagrangian
\be
{\cal L} = \partial x_1 \partial\bar{x}_1 + (x_1 \bar{x}_1)^{p-n+1}
\label{LG}
\ee
is incomplete, as can be seen from its having too large symmetry. It
is easy to show that a lower degree term with the correct symmetry 
has been omitted
\cite{I},
\be
\delta{\cal L} = (x_1 \bar{x}_1)^{p-n-1} \left(x_1^2 \bar{x}_2 +
\bar{x}_1^2 x_2 \right).
\label{LGadd}
\ee
Nevertheless, this Lagrangian is still incomplete. The total Lagrangian
can be found for $W_3$-models using methods of singularity theory \cite{II}.

When there is $N=2$ supersymmetry one can identify the fusion rules of a
subalgebra of primary fields, that of the chiral fields, 
with the fusion rules of the affine algebra $SU(n)_k$
\cite{Gepner}. Thus the critical superpotential turns out to be the
quasi-homogeneous part of the fusion potential. This potential, $\W$ say, is
generally complex but there is an associated bosonic potential given by
$V=|\partial_i\W |^2$, with the property that the extrema of $\W$
correspond to zero energy ground states of $V$. This should be the first
candidate of which one could think for the potential of 
the non supersymmetric $W$-models. However, we shall see in the next section
that it is not suitable. Nevertheless, we can still identify within the
operator algebra fields with the fusion rules
of the affine algebra $SU(n)_k$, modulo irrelevant fields. These fields
are the diagonal fields 
$\Phi(\l\mid\l)$. Relying on this fact, we may expect to be able to use
the known realization of $SU(n)_k$ fusion algebras in terms of orthogonal
polynomials \cite{Verlinde, Gepner} as well as in the the case with
$N=2$ supersymmetry.  

\section{Constructing the Landau potentials}

First of all, let us see that the bosonic potential $V=|\partial_i\W|^2$
of $N=2$ supersymmetric $W$-models is not a good candidate in the case
with no supersymmetry. The Gepner fusion potential $\W$ for $SU(n)_k$ is of
degree $k+n$. Hence the bosonic potential $V$ is of degree
$2(k+n-1)$. It does not match the degree of the potentials found before
(\ref{LG}), which is $2(p-n+1)=2(k+1)$ (except when $n=2$). We saw in a
number of cases \cite{I} that the latter potential suffices to produce
the correct number of minima under a suitable perturbation.
Thus the degree of the former potential being larger means
that one should not exclude in principle the existence of further
minima, though they would not have zero energy. 
The possible presence of extra minima would be nevertheless an
undesirable feature. 
The root of the problem is that the form of the unperturberd $N=2$
bosonic potentials 
does not constitute a bona fide real singularity (except for $n=2$).%
\footnote{This fact seems to have been overlooked in the literature.
Presumably, it is not crucial for the $N=2$ case, where one is
essentially interested in the chiral ring and hence the holomorphic
potential $\W$, which is a well defined complex singularity.}
In relation to it one could add another reason to discard the
$N=2$ bosonic potential: Let us assign degree $k$ to $x_k$ and $\bar{x}_k$,
$k\leq[n/2]$. When the perturbation is set to zero this potential
becomes inhomogeneous whereas the unperturbed potential given by
(\ref{LG}) and (\ref{LGadd}) is homogeneous.
We should look for a perturbed
potential whose (quasi-)homogeneous part agrees with (\ref{LG}) and
(\ref{LGadd}). 

Let us recall the reason why the bosonic potential has as zero-energy
ground states a set of points corresponding to the Weyl alcove of level $k$: 
They are the solutions of the equations $\partial_i\W = 0$
\cite{Gepner}. Indeed, one
can see that these equations imply the vanishing of the polynomials 
representing fields at level $k+1$. In the language of algebraic geometry
one can actually identify (as a category functor) 
a finitely generated algebra with
a set of points. Since we expect precisely the set of points above to be 
the minima of the perturbed potential for which we are looking, we must
also expect that 
an algebra equivalent to that of $SU(n)_k$ to appear. It does indeed appear as
the fusion algebra of diagonal fields. To better understand their physical
r{\^o}le we must bring about the connection with integrable IRF models. The 
ground states of these models are in correspondence with the points
of a graph, the mentioned Weyl alcove \cite{JMO}. 
It is natural to define a set of order parameters to
characterize these ground states. The first candidates are local state
probabilities, defined as the expectation value of the proyector onto a
definite ground state. They are the analog of the point set basis of
the algebra, that is, the basis that consists of functions that vanish in
all the points except one, for each of them \cite{Fulton, Gepner}.
However, it is preferable to take the linear
combinations $$\Phi^{(a)}({\bf r})=\sum_{\alpha} {\psi^{(a)}_{\alpha}\over
\psi^{(1)}_{\alpha}}\, \cal{P}_{\alpha}({\bf r}),$$
with $\psi^{(a)}_{\alpha}$ the eigenvectors of the adjacency matrix of
the graph and $\cal{P}_{\alpha}$ the local state probability of height
$\alpha$.  
Their continuum limit gives the diagonal fields, which
are hence understood as the order parameters \cite{Pasquier}. (One
should exclude the identity 
operator.) They distinguish between ground states. 

>From the discussion above we conclude that the conditions for the vanishing
of fields at level $k+1$ must play a decisive role in constructing the
potential. Furthermore, we would like the structure of the 
potential to show clearly that the ground states are the ones given by those
conditions. The $N=2$ bosonic potential fulfils these requirements but is not
suitable. Fortunately, there is another possibility: In place of 
the derivatives of the Gepner potential we may take the polynomials that
represent the fields at level $k+1$ and define%
\footnote{These polynomials are conjugates
of one another due to the reflection symmetry of the $A_{n-1}$ Dynkin
diagrams; hence the sum actually runs over half the weights
$\lambda^{(k+1)}$---or half plus one.}
\be
V = \sum_{\lambda^{(k+1)}} |P_\lambda (x_i)|^2.     \label{Lpot}
\ee
The first term occurs for $\l=(k+1)\o_1$ and has the form
\bea
|P_{(k+1)\o_1}(x_i)|^2 = P_{(k+1)\o_1}(x_i)\, P_{(k+1)\o_{n-1}}(x_i) =
\nonumber\\ 
(x_1^{k+1} - k\,x_2 x_1^{k-1} + \cdots) (\bar{x}_1^{k+1} - k\,\bar{x}_2
\bar{x}_1^{k-1} + \cdots) = \nonumber\\
(x_1\bar{x}_1)^{k+1} - k\,(x_1\bar{x}_1)^{k-1} 
\left(x_1^2 \bar{x}_2 + \bar{x}_1^2 x_2 \right) + \cdots . \label{term1}
\eea
Hence it agrees with (\ref{LG}) and (\ref{LGadd}). All the other terms
associated to the weights $\o_1$ and $\o_{n-1}$ only, namely,
$\l=(k+1-j)\o_1+j\o_{n-1},~j=1,\ldots,k$, yield a similar result (see
appendix A),
\be
|P_{\l}(x_i)|^2 = (x_1\bar{x}_1)^{k+1} - (k-1)\,(x_1\bar{x}_1)^{k-1} 
\left(x_1^2 \bar{x}_2 + \bar{x}_1^2 x_2 \right) + \cdots . \label{term1n-1}
\ee
The remaining terms in (\ref{Lpot}) have lower degree in
$x_1, \bar{x}_1$. This can be seen by noticing that the highest degree
monomial in $P_\lambda (x_i)$ is given by (\ref{diag}) with Dynkin labels
such that $\sum\mu_i = k+1$ at level $k+1$.
When $n=3$ then 
$x_{n-1} = x_2$ and the remaining terms do not appear, leaving just
(\ref{term1}) and (\ref{term1n-1}) with $x_2 = \bar{x}_1$. 

We may question
in general if all the terms in (\ref{Lpot}) are actually necessary. It
may well happen 
that the vanishing of a subset of the polynomials at level $k+1$
suffices to imply the vanishing of them all. We shall see next that this is
indeed so, although the subset of polynomials $\l=(k-j)\o_1+j\o_{n-1}$
is not the right one. We already know that the vanishing of the
derivatives of the Gepner potential, $$\partial_i\W =
P((k+n-i)\o_1),~i=1,\dots,n-1,$$ implies the vanishing of all the polynomials
at level $k+1$. One can prove this in a constructive way, making
combinations of $P((k+n-i)\o_1)$ to produce those polynomials. The first
one is naturally $P((k+1)\o_1)$. Multiplying it by $x_1$ we obtain
\be
x_1\,P((k+1)\o_1) = P((k+2)\o_1) + P(k\o_1+\o_2),   \label{poly}
\ee
whereby we deduce the vanishing of $P(k\o_1+\o_2)$. Continuing this
procedure we can deduce the vanishing of all polynomials of the form
$P(k\o_1+\o_i),~i=1,\dots,n-1,$ (see appendix B). From them one can proceed
with the rest of the polynomials at level $k+1$. Conversely, the
vanishing of $P(k\o_1+\o_i),~i=1,\dots,n-1,$ implies that of the
polynomials $P((k+n-i)\o_1)$. Thus we conclude that
the adequate subset of these polynomials to include in the potential is
precisely $P(k\o_1+\o_i),~i=1,\dots,n-1$. Therefore, the potential
(\ref{Lpot}) must include in its highest degree part terms like
$|x_1^k\,x_i|^2$, contradicting somehow the simple form (\ref{LG})
obtained by Zamolodchikov's method. However, their presence can be
 understood in this context as follows: The simple form (\ref{LG}) was
obtained from the field equation (\ref{fe}) but one can and must consider
other field equations coming from the product of $\epsilon^k$ with other
elementary fields,
\be
x_i\,(x_1 \bar{x}_1)^{p-n} = \partial^2 x_i.  \label{fei} 
\ee
They give rise to the desired terms. Henceforth we consider in
(\ref{Lpot}) only the terms $|P(k\o_1+\o_i)|^2,~i=1,\dots,n-1.$

Once we have the perturbed Landau potentials, it is straightforward to
obtain the critical potential: One just has to put a coupling constant
before the term $\epsilon^k = (x_1 \bar{x}_1)^k$ in the expansion of the
potential and its adequate
powers before the remaining terms according to dimensional analysis. Then
one lets the constant go to zero. These singular potentials are probably
equivalent to some of the real singularities listed in the literature,
though their complexity greatly hinders any analysis to that effect.

\section{Further properties of perturbed Landau potentials}

It is clear that the Landau potential (\ref{Lpot}) has the required minima
by construction. We may wonder about other properties, e.g., the other
extrema or the possible solitons connecting those minima. The existence
of maxima and saddle points is related with the presence of non-diagonal
relevant fields: In the algebraic-geometric picture presented in the
previous section extra points correspond to extra generators in the
algebra. The total algebra is defined by the relations $\partial_xV =
\partial_yV = 0$. The remaining generators can in general be obtained as
in \cite{II}; namely, one can use the relations to eliminate all but a
finite number of polynomials in the $x_i$. To just find their number it
may be sufficient to consider the solutions of the equations of motion
\cite{II}. However, the problem in general is very complicated. In any event,
these non-diagonal fields are not associated to any physical phase
transition (they are not order parameters), although they are
significant for the topology of the potential.

As regards the question about solitons, it has been conveniently solved
for theories with $N=2$ supersymmetry: The soliton spectrum is
constituted by elementary 
solitons that interpolate between neighbouring minima
and whose energy saturates the Bogomolny bound \cite{Fendley et al}. 
This structure nicely agrees with what one expects from the connection
with solvable lattice IRF models and affine Toda field theories
\cite{Fendley-Intri}. Moreover, the soliton $S$-matrices are tensor
products of $N=0$ $S$-matrices with a fixed $N=2$ $S$-matrix.
Therefore, we should expect similar properties for both $N=2$ and
$N=0$ solitons.

One can see that the argument in
\cite{Fendley et al} still works for the present bosonic potentials,
due to their close relation with those of theories with $N=2$
supersymmetry. The complete form of the bosonic $N=2$ potential includes
a real K{\"a}hler metric $g$, 
\be
V=(g^{-1})^{{\bm x}\bar{{\bm x}}}\,|\partial_{\bm x}\W|^2.   \label{cLpot}
\ee
The proof of the essential properties of $N=2$ solitons goes
through irrespective of the form of that metric. Now we can see that
the potential proposed in the previous section can be put in this
form. This is due to the fact that the polynomials $P(k\o_1+\o_i)$ are
linear combinations of derivatives of the Gepner potential $\W$
with $x_i$-dependent coefficients (\ref{poly})(see appendix B for the
complete expressions). Hence, the potential can be
written in the form 
(\ref{cLpot}) for some polynomial $g^{-1}$. The metric is not 
K{\"a}hler (there is no reason why it should be) but it does not matter in a
purely bosonic theory.
Furthermore, the determinant of $g^{-1}$ is 1 (appendix B), which implies
that $g$ is also polynomial. Hence, the kinetic term of the
Landau-Ginzburg Lagrangian, $g_{{\bm x}\bar{{\bm x}}}\,\partial{\bm
x}\partial\bar{{\bm x}}$, is equivalent 
to its basic part, $\partial{\bm x}\partial\bar{{\bm x}}$, modulo a
finite number of irrelevant fields. 

The peculiar properties of $N=2$ solitons are directly related with the
existence of the supersymmetries. Then it may seem rather unexpected to
find similar properties in a purely bosonic theory. Its ultimate reason
probably lies in the presence in these theories of quantum affine
symmetry. This powerful symmetry is the reason for their
integrability. Besides, it has been shown that it can be understood as
a sort of fractional supersymmetry \cite{Ahn}. Moreover, this
symmetry can form the basis of a generalized chiral-ring structure and
hence Landau-Ginzburg Lagrangians. A preliminary attempt in this
direction was presented for $SU(2)$ in
\cite{LeCl-Vafa}. The elementary field considered there was also the
most relevant field. The chiral algebra would correspond to the diagonal
fields here. However, the ensuing Lagrangians are the straightforward
generalization of the $N=2$ ones, which thence were formulated in terms of
pseudo-Grassman variables, and there was made no attempt to
construct from them the purely bosonic Landau-Ginzburg Lagrangians. 
One can speculate that the
Landau-Ginzburg Lagrangians constructed here are suitable candidates to
fit in a chiral-ring picture once generalized to $n>2$.

\section{Conclusion}

We have combined Zamolodchikov's method with known results for
perturbed $N=2$ potentials in order to 
construct perturbed $N=0$ potentials.
To make them agree we propose an alternative to the bosonic
$N=2$ potential which reduces its degree to the
correct value; namely, we propose to select a different set of vanishing
polynomials that also produce truncation at level $k$. 
Finally, it was shown that this modification only amounts to the
introduction of 
a non-trivial hermitian (though non-K{\"a}hler) metric in the bosonic
$N=2$ potential. Therefore, the essential
properties of these potentials, in particular as
regards to solitons, are preserved.

Our bosonic potentials coincide with the bosonic $N=2$ potentials for
the one-variable ($n=2$) case. Moreover, in the $n=3$ case the
unperturbed part of the potentials is that previously found in
\cite{II}. The potentials for $n>3$ constitute the natural
generalization of those previous results. Since these potentials 
represent $W$-models
perturbed by the least relevant field, one can produce the corresponding
multicritical second-order phase transition by turning off that perturbation.
Below the transition the coupling is negative and we have the full
unfolding of minima. Above the transition the coupling 
is positive and the perturbation $\epsilon^k=(x_1 \bar{x}_1)^{k}$ becomes
irrelevant, that is, turns into the new potential. 
Thus the coupling constant of this perturbation must be associated to 
the temperature, hence reproducing the phase transition between regimes
III and IV of the Jimbo {\em et al} IRF models \cite{JMO}.

The construction relies on the existence in the $W$-models of a
$SU(n)_k$ algebra of diagonal fields and hence of a Gepner
potential. According to results for $n=2$ \cite{LeCl-Vafa} this
algebra could be further 
given the structure of a chiral algebra in an analogous sense as for
$N=2$. This possibility would suggest that the Gepner potential itself
has a physical r{\^o}le in the bosonic theories and methods of
complex singularity theory similar to the ones used in $N=2$ (chiral
rings) might be also applicable to $N=0$. 

\vspace{1cm}
I acknowledge hospitality at the Department of Mathematics of The
University of Durham, where this work was started. I also thank
P.\ Bowcock, C.\ G{\'o}mez, P.\ Mansfield and G.\ Sierra for conversations. 

\appendix
\section{The polynomials in the $(\o_1,\o_{n-1})$ plane}
Let us find the form of the polynomials
$P(\l=(k+1-j)\o_1+j\o_{n-1}),~j=1,\ldots,k$. According to \cite{Gepner}
they are given by the Giambelli formula
\bea
P((k+1-j)\o_1+j\o_{n-1}) =
[\underbrace{1,\ldots,1}_{k+1-j},\underbrace{n-1,\ldots,n-1}_{j}] =
\nonumber  \\
\nonumber  \\
\left|\begin{array}{cccccccccc} 
x_1 & 1&0&&\cdots&&\cdots&&&0\\
x_2 & \ddots & \ddots& \ddots\\ 
0&\ddots & \ddots &\ddots&\ddots\\
&\ddots&\ddots&\ddots& 1 & 0 &&&&\vdots\\ 
\vdots&&\ddots& x_2 & x_1 & 1 &0\\ 
&&&0& 1 & x_{n-1} &x_{n-2}&\ddots&&\vdots\\
\vdots&&&&0& 1 & \ddots&\ddots&\ddots\\ 
&&&&&\ddots&\ddots& \ddots& \ddots&0  \\
&&&&&&\ddots& \ddots &\ddots &x_{n-2} \\ 
0&&&\cdots&&\cdots&&0& 1 & x_{n-1}\end{array}\right|
=  \nonumber  \\
~~\nonumber  \\
x_1^{k+1-j}x_{n-1}^j - (k-j)x_1^{k-j-1}x_2\,x_{n-1}^j -
(j-1)x_1^{k+1-j}x_{n-2}\,x_{n-1}^{j-2} + \cdots.
\eea
Hence
\be
|P((k+1-j)\o_1+j\o_{n-1})|^2 = (x_1\bar{x}_1)^{k+1} -
(k-1)\,(x_1\bar{x}_1)^{k-1} \left(x_1^2 \bar{x}_2 + \bar{x}_1^2 x_2
\right) + \cdots .
\ee

\section{Relation between $P(k\o_1+\o_i)$ and $\partial_j\W$}
Here we shall show how to express the polynomials
$P(k\o_1+\o_i),~i=1,\dots,n-1$, in terms of the polynomials $\partial_j\W =
P((k+n-j)\o_1),~j=1,\dots,n-1.$ We need to generalize eq.\ (\ref{poly})
to $i>2$. For this we take $P((k+j)\o_1)$ ($j$ immaterial) and multiply
it by an arbitrary $x_i,~i=1,\dots,n-1$. In this product shall appear
any $P(\l)$ such that $\l$ belongs to the Clebsch-Gordan decomposition
of $(k+j)\o_1 \oplus \o_i$ (in an obvious notation), namely, $\l=(k+j)\o_1
+ \o_i$ and $\l=(k+j)\o_1$ plus a Weyl transform of $\o_i$. All these
transforms except the first take $(k+j)\o_1$ out of the fundamental Weyl
dominion and therefore yield a null result. The first transform is given
by $\o_i=\e_1+\cdots+\e_i \rightarrow\e_2+\cdots+\e_{i+1}$ with
$\left\{\e_i\right\}_{i=1}^n$ the projection of an orthonormal basis of
\IR$^n$, $\left\{{\bf e}_i\right\}_{i=1}^n$, orthogonal to $\sum_{i=1}^n{\bf
e}_i$. Since 
$\e_2+\cdots+\e_{i+1}= -\o_1+\o_{i+1}$ we finally find that
\be
x_i\,P((k+j)\o_1) = P((k+j)\o_1+\o_i) + P((k+j-1)\o_1+\o_{i+1}).
\ee
If $i=j=1$ we have eq.\ (\ref{poly}). When $i+j=3$ we have two equations
and likewise onwards. Adding the $m-1$ equations for $i+j=m$ with
alternate signs, 
all $P((k+j-1)\o_1+\o_{i+1})$ cancel except the last,
$P(k\o_1+\o_{m})$. Solving for it for every $m=2, \dots, n-1$ we obtain
the set of relations 
\bea
P(k\o_1+\o_{m}) = x_{m-1}  P((k+1)\o_1) - x_{m-2} P((k+2)\o_1) + x_{m-3}
P((k+3)\o_1) - \nonumber  \\ \cdots \pm P((k+m)\o_1)
\eea
or, in matrix form,
\be
\left[\begin{array}{c} P((k+1)\o_1)\\P(k\o_1+\o_2)\\P(k\o_1+\o_{3})\\
\vdots \\P(k\o_1+\o_{n-1}) \end{array}\right] =
\left[\begin{array}{ccccc} 
1&0&&\cdots&0 \\ 
x_1&-1&\ddots&&\vdots \\
x_2&-x_1&1&\ddots&\vdots \\ 
\vdots&\vdots&&\ddots&0 \\
x_{n-2}&-x_{n-3}&\cdots&&\pm 1 \end{array}\right]
\left[\begin{array}{c} P((k+1)\o_1)\\P((k+2)\o_1)\\P((k+3)\o_1)\\ \vdots
\\P((k+n-1)\o_1)\end{array}\right].
\ee
If we call the transformation matrix $T$, it is immediate that $\det
T=\pm 1$. 

The potential (\ref{Lpot}) can thence be expressed as
\be
V = \sum_{\lambda^{(k+1)}} |P_\lambda (x_i)|^2 = \sum_{\mu\bar{\mu}}
T_{\lambda\mu} \overline{T_{\lambda\mu}} \,|P_\mu (x_i)|^2,
\ee
where $\l = k\o_1+\o_{m}$ and $\m = (k+m)\o_1$ ($m=1, \dots, n-1$). Therefore,
\be
V=(g^{-1})^{{\bm x}\bar{{\bm x}}}\,|\partial_{\bm x}\W|^2
\ee
with 
\be
(g^{-1})^{{\bm x}\bar{{\bm x}}} = T\,
\overline{T}\,\delta^{{\bm x}\bar{{\bm x}}}.
\ee
Note that $\det g^{-1} = 1$.

\end{document}